\documentclass[prd,aps,oamsmath,amssymb,nofootinbib,preprintnumbers]
{revtex4}
\usepackage{graphicx}
\usepackage{dcolumn}
\usepackage{bm}
\usepackage{amsmath}
\usepackage{amsthm}

\usepackage{amsfonts}
\usepackage{euscript,bbm}
\usepackage{ifthen}
\usepackage{psfrag}
\usepackage{slashed}
\usepackage{hyperref}

\def\ls{\mathrel{\lower4pt\vbox{\lineskip=0pt\baselineskip=0pt
           \hbox{$<$}\hbox{$\sim$}}}}
\def\gs{\mathrel{\lower4pt\vbox{\lineskip=0pt\baselineskip=0pt
           \hbox{$>$}\hbox{$\sim$}}}}
\def\drawbox#1#2{\hrule height#2pt

\hbox{\vrule width#2pt height#1pt \kern#1pt
              \vrule width#2pt}
              \hrule height#2pt}

\def\Asym#1#2{\vcenter{\vbox{\drawbox{#1}{#2}
              \kern-#2pt       
              \drawbox{#1}{#2}}}}

\newcommand{\be}{\begin{equation}}
\newcommand{\ee}{\end{equation}}
\newcommand{\bea}{\begin{eqnarray}}
\newcommand{\eea}{\end{eqnarray}}
\providecommand{\e}[1]{\ensuremath{\times 10^{#1}}}

\newcommand{\slep}{\ensuremath{\tilde{l}_{L}}}
\newcommand{\sneu}{\ensuremath{\tilde{\nu}}}
\newcommand{\neu}[1]{\ensuremath{\tilde{\chi}_{#1}^0}}

\newcommand{\chpm}[1]{\ensuremath{\tilde{\chi}_{#1}^{\pm}}}

\newcommand{\gsim}{\lower.7ex\hbox{$\;\stackrel{\textstyle>}{\sim}\;$}}
\newcommand{\lsim}{\lower.7ex\hbox{$\;\stackrel{\textstyle<}{\sim}\;$}}

\newcommand{\ttbar}{t \bar{t}}

\newcommand{\met} {{E\!\!\!\!/_{T}}}

\newcommand{ \pythia } {{\tt PYTHIA}}

\newcommand{ \pgs }    {{\tt PGS4}}

\newcommand{ \madgraph } {{\tt MADGRAPH5}}

\begin{document}

\begin{flushright}
MIFPA-14-33
\end{flushright}

%
\title{Probing Compressed Sleptons at the LHC using Vector Boson Fusion Processes}


\author{Bhaskar Dutta$^{1}$}
\author{Tathagata Ghosh$^{1}$}
\author{Alfredo Gurrola$^{2}$}
\author{Will Johns$^{2}$}
\author{Teruki Kamon$^{1,3}$}
\author{Paul Sheldon$^{2}$}
\author{Kuver Sinha$^{4}$}
\author{Kechen Wang$^{1,5}$}
\author{Sean Wu$^{1}$}

\affiliation{$^{1}$~Mitchell Institute for Fundamental Physics and Astronomy, \\
Department of Physics and Astronomy, Texas A\&M University, College Station, TX 77843-4242, USA \\
$^{2}$~Department of Physics and Astronomy, Vanderbilt University, Nashville, TN 37235, USA\\
$^{3}$~Department of Physics, Kyungpook National University, Daegu 702-701, South Korea \\
$^{4}$~Department of Physics, Syracuse University, Syracuse, NY 13244, USA \\
$^{5}$~Center for Future High Energy Physics, Institute of High Energy Physics, Chinese Academy of Sciences,
Beijing 100049, China
}

\begin{abstract}

The vector boson fusion (VBF) topology at the Large Hadron Collider at 14 TeV provides an opportunity to search for new physics. A feasibility study for the search of  sleptons in a compressed mass spectra scenario is presented in the final state of two jets, one or two low $p_{T}$ non-resonant leptons, and missing energy. The presence of the VBF tagged jets and missing  energy are effective in reducing Standard Model backgrounds. Using smuon production with a mass difference between $\slep$ and $\neu{1}$ of 5-15 GeV, the significance of observing the signal events is found to be\\ $\sim$ 3-6$\sigma$ for $m_{\tilde{l}}$=115-135 GeV, considering an integrated luminosity of 3000 fb$^{-1}$.


\end{abstract}

\maketitle


\section{Introduction} Searches for supersymmetry (SUSY) at the Large Hadron Collider (LHC) have produced impressive constraints on colored superpartners. For comparable masses, the exclusion limits on squarks ($\tilde{q}$) and gluinos ($\tilde{g}$) are approximately $1.5$ TeV at $95\%$ confidence level with $20$ fb$^{-1}$ of integrated luminosity \cite{:2012rz, Aad:2012hm,cmssusy, :2012mfa}.

On the other hand, searches for charginos ($\chpm{1}$), neutralinos ($\neu{2}$), and sleptons ($\slep \equiv \tilde{e}_L, \tilde{\mu}_L$) through direct electroweak production face the difficulty that their production cross-sections are much lower, resulting in smaller exclusion bounds. Directly produced sleptons have been probed at both ATLAS \cite{ATLASSlep} and CMS \cite{CMSSlep1}, in final states containing opposite-sign same-flavor non-resonant dileptons and missing transverse energy ($\met$). The decay chain is $pp \rightarrow \slep \slep^* \rightarrow l^+ l^- \neu{1} \neu{1}$, with $Br(\slep \rightarrow l^- \neu{1}) = 1$. Constraints for right slepton masses, have also been set by these studies.

The mass separation $\Delta M = m_{\slep}-m_{\neu{1}}$ is an important factor in the resulting exclusion plots from both experiments. The exclusion limits are given on the $m_{\slep}$-$m_{\neu{1}}$ plane, and depend on $\Delta M$. The mass reach with $m_{\tilde{\chi}_{1}^{0}} = 0$ GeV is $m_{\tilde{l}_{L}} \sim 280$ GeV and 330 GeV for CMS and ATLAS respectively. In the CMS studies, for $m_{\slep} \sim$ 110-200 GeV, the excluded region has $\Delta M \sim 110$ GeV. In the ATLAS study,  the exclusion limits reach $m_{\tilde{l}} \sim 250$ GeV with $\Delta M \, \sim \, 100$ GeV, after which $\Delta M$ increases. For the right-sleptons, the mass reach with $m_{\tilde{\chi}_{1}^{0}} = 0$ GeV is $m_{\tilde{l}_{R}} \sim 180$ GeV and 250 GeV for CMS and ATLAS respectively.

Compressed spectra with smaller $\Delta M$ may have eluded these probes, and are important for a variety of theoretical reasons. For example, for a Bino-like $\neu{1}$ dark matter (DM) candidate, the annihilation cross-section is usually too low, and needs to be considerably enhanced with coannihilation processes (typically requiring $\Delta M \, \sim \,$ 5-15 GeV for slepton coannihilation~\cite{Griest:1990kh,Coann}) to obtain the relic density observed by WMAP~\cite{WMAP}. Also, the annihilation diagrams, even when coannihilation is absent, may involve sleptons in the t-channel. On the other hand, the main supersymmetric contributions to the muon $g-2$ are dominated by chargino-sneutrino and neutralino-smuon loop diagrams with not necessarily very large mass gaps. Furthermore, the BNL measured excess~\cite{BNL} for the anomalous magnetic moment of the muon is about 3.6$\sigma$ (2.4$\sigma$) using $e^{+}e^{-}$ ($\tau$) data~\cite{Davier:2013wwg} and can be explained with a SUSY mass spectra containing $\mathcal{O}(100)$ GeV neutralinos and sleptons (we refer to \cite{Endo:2013bba} for a recent summary). 

In this paper, we propose search strategies for $\slep$ pairs using the vector boson fusion (VBF) topology. The VBF topology has been used by the authors recently to probe the supersymmetric electroweak sector. In \cite{Dutta:2012xe}, the $\chpm{1}$-$\neu{2}$ system (where it is mostly the charged and neutral Wino) has been studied, while direct production of $\neu{1}$ dark matter by VBF processes has been proposed in \cite{Delannoy:2013ata}. Previously, sleptons were studied in the context of the LHC ~\cite{sleptonold}. As shown in these references, the requirement of two energetic jets in the forward region, in opposite hemispheres, and with large dijet invariant mass is very effective in reducing standard model (SM) backgrounds. Further, the VBF topologies naturally give rise to larger $\met$ since the momentum of the particles produced in the slepton system must balance the high $p_{T}$ of the scattered partons, which is of significant experimental importance as it allows for an additional handle to trigger on compressed spectra that is typically characterized by low $\met$ in non-VBF searches. Thus, in the compressed scenario, the $\neu{1}$ resulting from the $\slep$ decay carries significant $\met$, providing better control of the SM background. We develop our VBF search strategy to be particularly suited to probes in the region of small $\Delta M$, in particular focusing on the use of soft leptons to discriminate against the SM background. 

The structure of the paper is as follows. In Section \ref{searchstrategy} an outline of the search strategy is given, followed by results in Section \ref{results}. Conclusions are given in Section \ref{conclusion}.

\section{Search Strategy}\label{searchstrategy}
To probe $\slep$ production, the following processes are investigated: $pp \rightarrow \slep \, \slep^{*} \, jj, \, \slep \, \sneu^* \, (\slep^* \sneu)jj$. We note that the mass splitting between $\slep$ and $\sneu$ is $m^2_{\slep} - m^2_{\sneu} = - \frac{1}{2} m^2_{Z} \cos{2\beta}$~\cite{Barger:1993gh} and thus $m_{\slep}> m_{\sneu}$. Therefore, along with $\slep$ pair production we also consider $\slep \sneu^*$ ($\slep^* \sneu$) production.  Two separate studies were performed in the final states of $2j \, + \, 2l \, + \, \met $ (which targets $\slep \, \slep^*$ production), and $2j \, + \, 1l \, + \, \met $ (which targets both $\slep \, \slep^*$ and $\slep \, \sneu^*$ ($\slep^* \sneu$) production). In order to avoid having $\sneu$ as the lightest SUSY particle and consequently as the DM candidate, a scenario ruled out by a combination of relic density and direct detection constraints~\cite{SneutrinoDM}, $\tan{\beta}$ is kept small [$\sim \mathcal{O}(1)$] so that $m_{\sneu} > m_{\neu{1}}$. Furthermore, to satisfy the muon $g-2$ constraint, $m_{\neu{i}},m_{\chpm{j}}$ (where, $i=2,3,4$ and $j=1,2$) are set to $\lesssim 1$ TeV, but it does not influence our analysis.

Several $\slep$ masses in the range $115-135$ GeV  are chosen for the study at $\sqrt{s} = 14$ TeV. For each $m_{\slep}$, $\Delta M$ is varied between 5 and 15 GeV (except for the $m_{\slep}=135$ GeV points, where $\Delta M$ between 5 and 25 GeV is considered). The $\neu{1}$ is purely Bino and the slepton is mostly $\slep$. The decay mode of $\slep$ is $\slep \, \rightarrow l \neu{1}$ with $100\%$ branching ratio. The rest of the spectrum is assumed to be much heavier. The signal samples are generated at $\mathcal{O}(\alpha^{4}_{EW} \alpha^{0}_s)$ and include 2-partons (exclusive) processes.~\footnote{For the benchmark point $(m_{\tilde{l}_{L}},m_{\tilde{\chi}_{1}^{0}})=(135$ GeV$, 120$ GeV$)$, we have also generated and analysed a sample up to $\mathcal{O}(\alpha^{4}_{EW} \alpha^{4}_S)$, which includes up to 3 partons (inclusive) processes, but did not find any significant increase in the signal rate and significance.}

The search strategy is based on three steps. First, we use the unique features of VBF processes to reduce non-VBF backgrounds by requiring large $\met$, $H_{T}$ 
and exactly two forward jets in opposite hemispheres and with large dijet invariant mass. Second, events containing additional leptons and/or b-tagged jets are rejected in order to reduce the contribution from WZ and $t\bar{t}$. Finally, the $p_{T}$ distributions of the muons are used to assess the presence of a signal above the SM background. We study the signal significance using two approaches: (1) a simple cut and count approach optimized for our benchmark scenario by requiring soft leptons; (2) a more general shaped based approach using a fit of the entire muon $p_{T}$ spectrum to search for an enhancement in the soft part of the spectrum.

Signal and background samples are generated with \madgraph \, \cite{Alwall:2011uj} followed by the parton showering and hadronization with \pythia \, \cite{Sjostrand:2006za} and the detector simulation using \pgs \, \cite{pgs}. We use CTEQ6L1~\cite{Pumplin:2002vw} parton distribution function.

\section{Results} \label{results}
 The SM backgrounds considered for this study are, $\ttbar \, + \,$ jets, $VV \, + \,$ jets, and $V \, + \,$ jets, where $V$ denotes $W, Z$. The $V \, + \,$ jets background is generated including up to 4-partons (inclusive), while the $\ttbar \, + \,$ jets calculation includes up to 3-partons (inclusive). Double-counting is avoided by using the MLM-scheme~\cite{Mangano:2006rw} for jet matching. The $VV \, + \,$ jets background is calculated at $\mathcal{O}(\alpha^{4}_{EW} \alpha^{0}_s)$ only and includes up to 2-partons (exclusive). Single top ($tW$, $tq$) and $\gamma^* \, + \,$ jets backgrounds are found to be insignificant for the following analyses. The Higgs background is negligible after applying the VBF cuts. For an inclusive Higgs sample of events equivalent to $\sim 3000$ fb$^{-1}$, we find no event passing after applying all cuts and so, the effective cross-section is 0 with an uncertainty of $3\times10^{-4}$ fb.
As a benchmark scenario we consider smuon production with $l=\mu$, but this study can  also be extended to the $l=e$ case. We use JETCLU-like~\cite{Abe:1991ui} cone algorithm as used by \pgs \, for jet reconstruction with cone radius=0.5. We further use $\Delta R = 0.3$ for jet isolation. The next-to-leading order QCD corrections to the VBF electroweak production of signal and background cross sections
have not been considered. The change in the signal~\cite{kfactor} and background~\cite{kfactorb} cross-sections due to the inclusion of the K factor is very
modest (at a few percent level) for VBF production.

\begin{table}[!htp] 
\caption{[$2 j + 2 \mu + \met $ study] Summary of the effective cross-section (fb) for the signal and main sources of background at LHC14 for the benchmark point $(m_{{\tilde{\mu}_L}}, m_{\neu{1}}) = (135,120)$ GeV. ``---'' indicates the background size is negligible. }
\label{2muBenchAndBgTable}
\begin{center}
\begin{tabular}{l c c c c c} 

\hline
\hline 
Selection & (135, 120) & $VV$ + jets & $t\bar{t}$ + jets & $W$ + jets & $Z$ + jets  \\  
  & [GeV] & [fb] & [fb] & [fb]  & [fb]  \\      

\hline 

Initial                 & 0.491 & 1.34$\e{3}$ & 7.03$\e{5}$ & 1.88$\e{8}$ & 5.56$\e{7}$ \\
exactly 2 j             & 0.265 & 2.08$\e{2}$ &   3.32$\e{4}$ &  2.38$\e{7}$ &  9.61$\e{6}$ \\
VBF topology selection  & 0.0327 &   4.29 &      35.4 &      4.68$\e{3}$ &     1.58$\e{3}$ \\
exactly 2 muon          & 0.0085 &   0.125 &       0.500 &         --  &          -- \\
$p_{T_{\mu_1}}+p_{T_{\mu_2}}<70$&0.0062 & 0.0126 &    0.0700 &        --     &       --   \\
$\met > 200$           & 0.0021 &   0.0021 &        --   &        --     &       --    \\
$H_T > 200$            & 0.0021 &   0.0020 &        --   &         --     &       --     \\

\hline
\hline
\end{tabular}
\end{center}
\end{table}

$\mathbf{2j \, + \, 2l \, + \, \met} $ \textbf{Study -} The selections for the $2j \, + \, 2l \, + \, \met$ study are as follows:
\begin{itemize}
\item exactly 2 j:\\ 
    (i) $b$-veto, where we assume $b$-tagging efficiency ($\epsilon_b$) of $70 \%$ and a fake rate ($f$) of $1 \%$ coming from $u,d,s,c,g$ jets. Both $\epsilon_b$ and $f$ are flat over $p_{T} > 30$ GeV for $|\eta| < 2.4$;\\
    (ii) select exactly 2 jets with $p_{T_j} > 30$ GeV and $|\eta_j| < 5$.
\item VBF topology selection:\\
    (i) $\eta_{j_1}\eta_{j_2} < 0$ and $M_{j_1j_2} > 600$ GeV;\\
    (ii) $|\eta_{j_{1,2}}| > 1.7$;\\    
    (iii) $\Delta\phi_{j_1j_2}<1.0$.
 \item exactly 2 muons: \\
    (i) select 2 muons, having opposite charges, with $p_{T_{\mu}} > 10$ GeV and $|\eta_{\mu}| < 2.5$;\\
    (ii) veto events with a loosely identified $e, \tau$;\\    
    (iii) $Z$-veto (i.e. reject events with $81$ GeV $< M_{\mu^{\pm}_{1}\mu^{\mp}_{2}} <$  $101$ GeV);\\
    (iv) Central $\mu$ selection with $\eta_j({\rm min}) < \eta_\mu < \eta_j({\rm max})$, where $\eta_j({\rm min})$ and $\eta_j({\rm max})$ are the minimal and maximal $\eta$s of two jets.\\
\end{itemize}

\begin{figure}[!htp]
\centering
\includegraphics[width=4.0in]{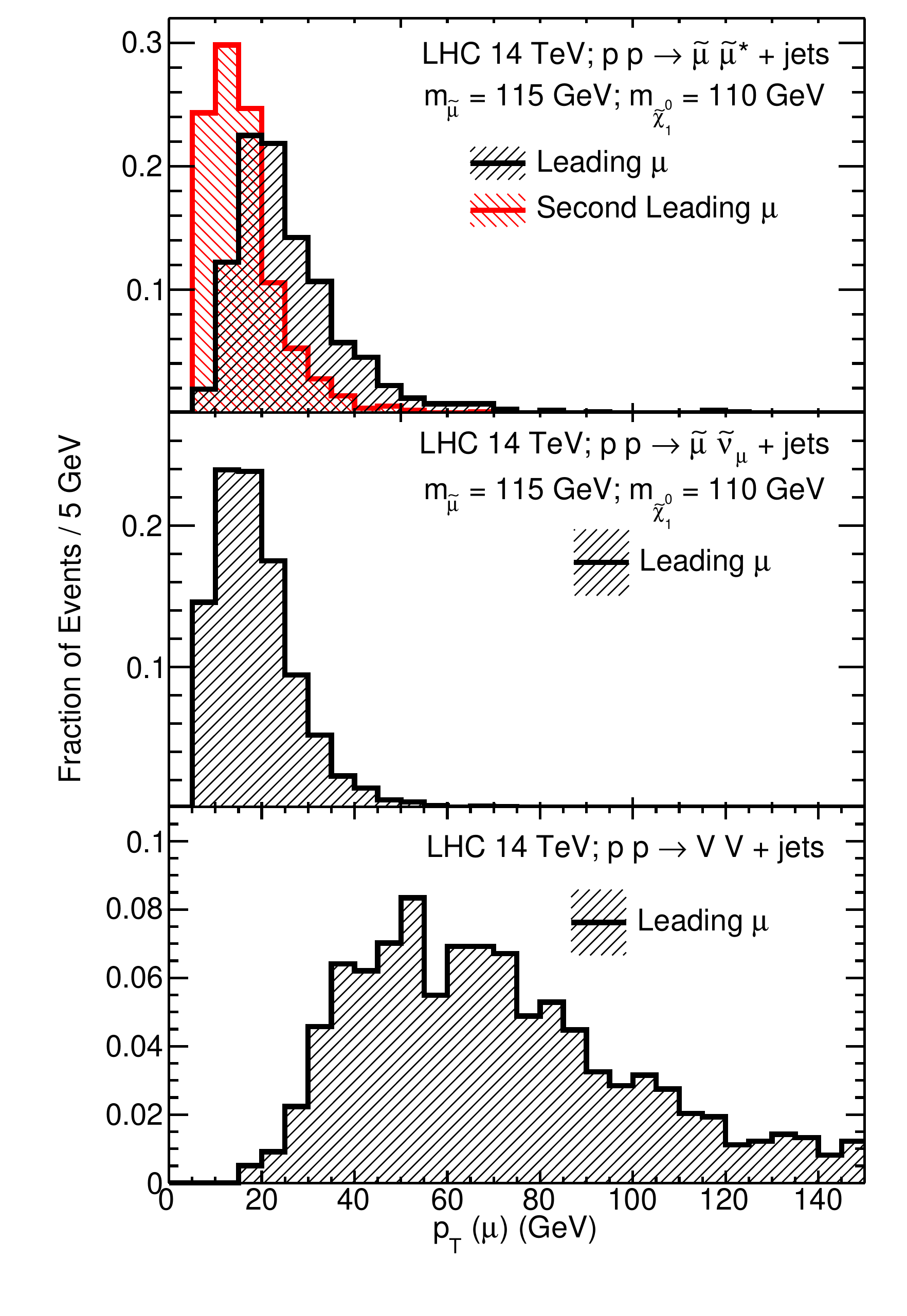}
\caption{Distribution of $p_{T_{\mu}}$ in the $\slep$ pair, $\slep \sneu^*$ ($\slep^* \sneu$) and $VV \, + \,$jets events for the benchmark point $(m_{\tilde{l}_{L}},m_{\tilde{\chi}_{1}^{0}})=(115$ GeV$, 110$ GeV$)$. All distributions shown in the figure are after VBF cuts, $\met$ and $H_T$ requirements but without the $p_{T_{\mu}}$ upper-bound cuts. This figure is a representative of the plots used for our shape based analysis.}
\label{muonpt}
\end{figure}

As shown in Fig.~\ref{muonpt}, the muons from the smuon decays are expected to be soft in the compressed spectra benchmark scenarios considered, with $p_{T} \sim \Delta M$.  For the cut and count (CC) approach, we take advantage of this characteristic by imposing an upper cut on the scalar sum of the $p_{T}$ of both muons ($p_{T_{\mu_{1}}}+p_{T_{\mu_{2}}} < 70$ GeV). Finally, we require $\met > 200$ GeV and $H_T > 200$ GeV, where $H_T$ is the scalar sum of the transverse momentum of all jets with $p_T>30$ GeV (including $b$ and $\tau$ jets). The cross-sections after each set of cuts are shown in Table~\ref{2muBenchAndBgTable}. Since the single top and $\gamma^* \, + \,$  jets backgrounds are negligible for this analysis, they are not presented in Table~\ref{2muBenchAndBgTable}.
The contribution from $\ttbar \, + \,$ jets vanishes after all the selections, but  $VV \, + \,$ jets survives with a background rate comparable to the expected signal rate. 

\begin{table}[!htp]
\caption{[$2 j + 1 \mu + \met $ study] Summary of the effective cross-section (fb) for the signal and main sources of background at LHC14 for the benchmark point $(m_{\tilde{\mu}_L}, m_{\neu{1}}) = (135,120)$ GeV. ``---'' indicates the background size is negligible.}
\label{1muBenchAndBgTable}
\begin{center}
\begin{tabular}{l c c c c c} 

\hline
\hline 
Selection & (135, 120) & $VV$ + jets & $t{\bar{t}}$ + jets & $W$ + jes & $Z$ + jets  \\  

 & [GeV] & [fb] & [fb] & [fb] & [fb] \\
      
\hline

Initial                & 0.953 & 1.34$\e{3}$ & 7.03$\e{5}$ & 1.88$\e{8}$ & 5.56$\e{7}$ \\
exactly 2 j            & 0.514 & 2.08$\e{2}$ &   3.32$\e{4}$ &  2.38$\e{7}$ &  9.61$\e{6}$ \\
VBF topology selection & 0.0240 &   2.88 &       7.27 &      1.08$\e{3}$ &      2.85$\e{2}$ \\
exactly 1 muon         & 0.0130 &   0.483 &       1.31 &       1.54$\e{2}$ &       18.4 \\
$p_{T_{\mu_1}}<30$     & 0.0075 &   0.0690 &       0.232 &        51.4 &        9.18 \\
$\met > 200$           & 0.0040 &   0.0259 &       0.0770 &          --  &         --  \\
$H_T > 250$            & 0.0029 &   0.0189 &        --  &          --  &         --   \\

\hline
\hline

\end{tabular}
\end{center}
\end{table}

 $\mathbf{2j \, + \, 1 \, \, l \, + \, \met }$ \textbf{Study -} All other selections being equal, the channel containing exactly one muon suffers from larger background rates with respect to the two muon channel.  However, the backgrounds are reduced to manageable levels with more stringent VBF topological selections. The selections for the $2j \, + \, 1\, \, \mu \, + \, \met$ study are as follows:

\begin{itemize}
\item exactly 2 j:\\ 
    (i) $b$-veto (same as in the $2j \, + \, 2l \, + \, \met$ study);\\
    (ii) select exactly 2 jets with $p_{T_j} > 30$ GeV and $|\eta_j| < 5$.
\item VBF topology selection:\\
    (i) $\eta_{j_1}\eta_{j_2} < 0$ and $M_{j_1j_2} > 1100$ GeV ;\\
    (ii) $|\eta_{j_{1,2}}| > 1.8$;\\    
    (iii) $\Delta\phi_{j_1j_2}<1.0$.
\item1  muon:\\
    (i) select events containing exactly 1 muon with $p_{T_{\mu}} > 10$ GeV and $|\eta_{\mu}| < 2.5$;\\
    (ii) veto events with a loosely identified $e$ or $\tau$;\\
    (iii) Central $\mu$ selection with $\eta_j({\rm min}) < \eta_\mu < \eta_j({\rm max})$, where $\eta_j({\rm min})$ and $\eta_j({\rm max})$ are the minimal and maximal $\eta$s of two jets.\\
 \end{itemize}   

As discussed above, the muons from the smuon decays are expected to be soft. Thus for the CC approach, we impose an upper threshold on the transverse momentum of the muon ($p_{T_{\mu_{1}}} <30$ GeV). Similarly, $\met > 200$ GeV and $H_T > 250$ GeV are found useful discriminants in this channel. The cross-sections after each set of cuts are shown in Table~\ref{1muBenchAndBgTable}.

{\it {\bf Significance -}} The significances $S / \sqrt{S+B}$ for the CC approach outlined in Tables I and II, where $S$ and $B$ are the signal and background yields respectively, are shown in Table III. As mentioned, we also perform a shape based analysis~\footnote{For the shape based analysis, we used distributions of $p_{T_{\mu}}$, which are generated after imposing all the cuts except the $p_{T_{\mu}}$ upper-bound cuts, namely, $p_{T_{\mu_1}}+p_{T_{\mu_2}}<70$ for the 2-muon channel and $p_{T_{\mu_1}}<30$ for the 1-muon channel. One such distribution is presented in Fig~\ref{muonpt} for the benchmark point $(m_{\tilde{l}_{L}},m_{\tilde{\chi}_{1}^{0}})=(115$ GeV$, 110$ GeV$)$.} of the $p_{T}(\mu)$ and $p_{T}(\mu_{1}) + p_{T}(\mu_{2})$ distributions in the 1-muon and 2-muon channels, respectively, using a binned likelihood following the test statistic based on the profile likelihood ratio. A local p-value is calculated as the probability under a background only hypothesis to obtain a value of the test statistic as large as that obtained with a signal plus background hypothesis. The significance $z$ is then determined as the value at which the integral of a Gaussian between $z$ and $\infty$ results in a value equal to the local p-value. In Table~\ref{sigsAllSAndBgTable}, we show the significances for exactly 1-muon and exactly 2-muon final states and the combined significances, using the joint likelihood, for these two channels.
We find that the combined significance is $\gtrsim$ 4$\sigma$ considering 3000 fb$^{-1}$ luminosity for $\Delta M$=5-15 GeV and $m_{\tilde{l}}$=115-125 GeV. For $m_{\tilde{l}}$=135 GeV, we find that the significance is $\gtrsim$ 3$\sigma$ for $\Delta M$=10-15 GeV. The significance becomes smaller for larger $\Delta M$ since the emitted muon from the $\slep$ decay has larger $p_T$ which makes it more difficult to discriminate signal from the $VV$ + jets background.


From Table~\ref{2muBenchAndBgTable} and~\ref{1muBenchAndBgTable}, it might appear to the reader that the 1-muon channel is insignificant compared to the 2-muon channel, owing to it's smaller $S/B$ ratio. However, a closer look at Table~\ref{sigsAllSAndBgTable} will reveal that with decreasing $\Delta M$, 1-muon channel becomes increasingly more important and it becomes the dominant channel for $\Delta M \sim 5$ GeV. This is due to the fact that for smaller $\Delta M$, the muons coming from the slepton decays will become even more soft, and it will become increasingly difficult to detect both muons in the 2-muon channel.

Monojet searches (one boosted high $p_{T}$ jet plus missing energy) have been conducted by the LHC experiments~\cite{Khachatryan:2014rra,ATLAS:2012zim,ATLAS:2012ky} as an effective probe for compressed spectra. It is interesting to compare the sensitivity of the proposed VBF searches to compressed slepton production with that of monojet. We find the monojet analysis does not provide sensitivity in these compressed slepton scenarios due to the small cross-sections. For this comparison we consider a benchmark point $(m_{\tilde{l}_{L}},m_{\tilde{\chi}_{1}^{0}})=(135$ GeV$, 120$ GeV$)$ and apply the selections outlined in the 14 TeV projection analysis~\cite{Schwaller:2013baa}. The combined signal production cross-section of $\tilde{\mu}_{L}\tilde{\mu}_{L}^{*}$ and $\tilde{\mu}_{L}\tilde{\nu}^{*}$ + jets (up to 3 partons) is 486.5 fb, while the dominant backgrounds are $V$ + jets~\cite{Baer:2014cua,Han:2013usa}. Following the 14 TeV projection analysis of Ref.~\cite{Schwaller:2013baa}, we obtain less than $1\sigma$ significance at 3000 fb$^{-1}$ of integrated luminosity for the aforementioned benchmark point.

\begin{table}[!htp]
\caption{Summary of the effective cross-section (fb) and significances, with 3000 fb$^{-1}$ after all cuts for different SUSY points at LHC14. The effective cross-section of total standard model background after all cuts is 0.0020 fb for ``exactly 2-muon final state analysis'', and 0.0189 fb for ``exactly 1-muon final state analysis''. The significances presented are calculated by means of both ``cut and count (CC)'' and ``shape analysis'' methods. }
\label{sigsAllSAndBgTable}
\begin{center}
\begin{tabular}{c c c | c c c | c c c |c c c c} 

\hline
\hline 
{$\Delta M$} & {$m_{\tilde{l}}$} & {$m_{\tilde{\chi}_1^0}$} & \multicolumn{3}{c|}{2-muon final state} & \multicolumn{3}{c|}{1-muon final state} & \multicolumn{4}{c} {Combined} \\
\cline{4-9}
 &  &  & Cross-section & Significance & Significance & Cross-section & Significance & Significance &  \multicolumn{4}{c} {Significance}    \\
 
 
 \multicolumn{3}{c|}{[GeV]} & [fb] & CC & Shape & [fb] & CC & Shape & & CC & & Shape  \\
\hline 

25 & 135 & 110 & 0.0014 & 1.3 & 1.8 & 0.0021 & 0.8 & 1.3 & & 1.6 & & 2.3 \\
15 & 135 & 120 & 0.0021 & 2.1 & 2.6 & 0.0029 & 1.0 & 1.5 & & 2.5 & & 3.2 \\
10 & 135 & 125 & 0.0019 & 2.1 & 2.9 & 0.0044 & 1.8 & 2.9 & & 2.9 & & 4.5 \\
 5 & 135 & 130 & 0.0004 & 0.3 & 0.5 & 0.0036 & 1.5 & 2.2 & & 1.5 & & 2.1 \\
\hline

15 & 125 & 110 & 0.0024 & 2.4 & 3.1 & 0.0035 & 1.3 & 1.8 & & 3.0 & & 3.8 \\
10 & 125 & 115 & 0.0018 & 2.0 & 2.8 & 0.0043 & 1.8 & 2.8 & & 2.9 & & 4.8 \\
 5 & 125 & 120 & 0.0006 & 0.6 & 1.0 & 0.0046 & 1.9 & 4.1 & & 2.1 & & 3.9 \\
 \hline
15 & 115 & 100 & 0.0027 & 2.8 & 4.1 & 0.0043 & 1.6 & 1.8 & & 3.5 & & 4.6 \\
10 & 115 & 105 & 0.0021 & 2.3 & 3.4 & 0.0050 & 2.0 & 3.2 & & 3.3 & & 5.1 \\
 5 & 115 & 110 & 0.0007 & 0.6 & 1.1 & 0.0058 & 2.4 & 4.1 & & 2.5 & & 4.0 \\

\hline
\hline


\end{tabular}
\end{center}\end{table}

\section{Conclusion} \label{conclusion}The main result of this paper is that the VBF topology provides a feasible strategy to search for sleptons in the case where the mass separation with the lightest neutralino is $\sim$ 5-25 GeV. The mass range is of great interest  phenomenologically and where exclusion bounds in non-VBF studies are difficult to obtain. There is no current or projected constraint for the 14 TeV LHC run available for this region. Two separate studies were performed in the final states of $2j \, + \, 2l \, + \, \met $ (which targets $\slep \, \slep^*$ production), and $2j \, + \, 1l \, + \, \met $ (which targets both $\slep \, \slep^*$ and $\slep \, \sneu^*$ ($\slep^* \sneu$) production). 
 As a benchmark scenario we consider smuon production with $\Delta M$=5-15 GeV and arrive at a combined significance of $\sim$ 4-6$\sigma$ considering 3000 fb$^{-1}$ luminosity for $m_{\tilde{l}}$=115-125 GeV. For $m_{\tilde{l}}$=135 GeV, the significance is $\gtrsim$ 3$\sigma$ for $\Delta M$=10-15 GeV. 
 If the analysis were repeated with $p_T > 5$ GeV for muons, the signal acceptance for the $\Delta M = 5$ GeV scenario should increase but a background estimate for soft muons in high pile-up condition, expected in high-luminosity LHC, will be a challenging task. 


{\it {\bf Acknowledgements -}} 
We would like to thank Xerxes Tata for helpful discussions. This work is supported in part by DOE Grant No. DE-FG02-13ER42020, NSF Award PHY-1206044, and by the World Class University (WCU) project through the National Research Foundation (NRF) of Korea funded by the Ministry of Education, Science, and Technology (Grant No. R32-2008-000-20001-0). T.K. is also supported in part by Qatar National Research Fund under project NPRP 5-464-1-080. K.S. is supported by NASA Astrophysics Theory Grant NNH12ZDA001N.

\end{document}